\documentclass[aps,pra,twocolumn,showpacs]{revtex4}
\usepackage{graphicx}
\usepackage{dcolumn}
\usepackage{bm}

\usepackage{subfigure}
\usepackage{ragged2e}
\usepackage{xcolor}

\begin{document}



\title{Magnon-polaron formation in XXZ quantum Heisenberg chains}

\author{D. Morais, F. A. B. F. de Moura  W. S. Dias}
\address{Instituto de F\' isica, Universidade Federal de Alagoas, 57072-900 Macei\' o, Alagoas, Brazil}

\begin{abstract}
We study the formation of magnon-polaron excitations and the consequences of different time scales between the magnon and lattice dynamics. The spin-spin interactions along the 1D lattice are ruled by a Heisenberg Hamiltonian in the anisotropic form XXZ, in which each spin exhibits a vibrational degree of freedom around its equilibrium position. By considering a magnetoelastic coupling as a linear function of the relative displacement between nearest-neighbor spins, results provide an original framework for achieving a hybridized state of magnon-polaron. Such state is characterized by high cooperation between the underlying excitations, where the traveling or stationary formation of magnon-polaron depends on the effective magnetoelastic coupling. A systematic investigation reveals the critical amount of the magnon-lattice interaction ($\chi_c$) necessary to emergence of the stationary magnon-polaron quasi-particle. Different characteristic time scales of the magnon and the vibrational dynamics unveiled the threshold between the two regimes, as well as a limiting value of critical magnetoelastic interaction, above which the magnon velocity no longer interferes at the critical magnetoelastic coupling capable of inducing the stationary regime. 
\end{abstract}

\maketitle

\section{Introduction}
\label{sec:int}

Collective excitations in magnetic ordering of a material result in the well-known spin waves, whose the quanta are called magnon. Prospects of using magnons for wave-based computation and data transfer have been studied~\cite{Chumak2014,Chumak2015,DEMIDOV20171,PhysRevX.5.041049,Grundler2016}, where recent advances such as magnon transistors~\cite{Chumak2014}, spin-Hall oscillators~\cite{DEMIDOV20171} and spin-wave diodes~\cite{PhysRevX.5.041049} are reported. Although magnonics is a promising approach~\cite{Krawczyk_2014,LENK2011107}, designing and controlling such quantum processes for long-time dynamics primarily requires understanding the role played by different ingredients.

Lattice vibrations are a key condensed matter topic, which have been remarkable effect to charge transport in polymers~\cite{SSH1979} and molecular crystals~\cite{holstein1959a,holstein1959b}. Although the first studies of magnetoelastic coupling of magnons and lattice vibrations have been developed some time ago~\cite{RevModPhys.25.233,PhysRev.110.836,kaganov1959}, such aspect has been actively studied on both experimental ~\cite{dai2000,reutler2002,man2017,LOU201837,milosavljevic2018,holanda2018,PhysRevB.102.144438,Berk2019,PRBmagnonphononjunho2020,naturedezem2020} and theoretical framework~\cite{li2017,PhysRevB.89.184413,PhysRevB.102.104411,PhysRevLett.124.147204}. Studies of magnetoelastic coupling in ferromagnetic manganese  perovskites have shown that the spin-wave softening and broadening are related to nominal intersection of the magnon and optical phonon modes~\cite{dai2000}. Inelastic neutron scattering was used to study low-energy ferromagnetic magnons and acoustic phonons in the ferrimagnetic insulators, whose results provided evidence for the presence of magnon-phonon~\cite{man2017}. Signatures of interaction between spin and phonon have been reported for magneto-thermal transport measurements in p-doped Si~\cite{LOU201837}, as well as layered semiconducting ferromagnetic compound CrSiTe$_3$ explored by raman scattering experiments~\cite{milosavljevic2018}. Probing the interaction between magnon and lattice vibrations revealed a conversion of spin wave-packets into an elastic wave-packets in films of a ferrimagnetic insulator under non-uniform  magnetic fields~\cite{holanda2018}. The  magnon-phonon  coupling  and  the  formation of an optically  excited magnon-polaron with high cooperation were described for a metallic ferromagnet with a nanoscale periodic surface pattern, where symmetries of the localized magnon and phonon states have been reported decisive for the hybridized state formation~\cite{PhysRevB.102.144438}. A strong coupling between magnons and phonons has been detected in the thermal conductivity of antiferromagnet $Cu_3TeO_6$~\cite{PRBmagnonphononjunho2020}. Magnetoelastic coupling has been theoretically studied to excite spin waves in magneto strictive films through surface acoustic waves on piezoelectric substrates, in which driven spin waves were able to propagate up to $1200~\mu m$~\cite{li2017}. A theoretical study shows the possibility of generating magnon-phonon coupling through a variable magnetic field in space, whose interaction depends on the strength of the magnetic field gradient~\cite{PhysRevB.102.104411}.   
The problem of a two-dimensional antiferromagnet  at the presence of
magnetoelastic coupling it was investigated in ref. \cite{PhysRevLett.124.147204} . The authors demonstrate that  the  magnon and phonon bands are  hybridized due to the magnon-phonon coupling, with the properties of the magnon-phonon excitations suggesting a nontrivial $SU(3)$ topology.

Another exciting experimental development involving interaction between magnons and lattice vibrations were the spin Seebeck effect~\cite{Uchida2008,Uchida2010,doi:10.1063/1.3529944} and the bottleneck accumulation of hybrid magnetoelastic bosons~\cite{PhysRevLett.118.237201}. The first one describes a spin current that appear in magnetic metal systems under effect of a thermal gradient, whose effect is enormously enhanced by non-equilibrium phonons~\cite{doi:10.1063/1.3529944}.  The bottleneck accumulation phenomenon for a magnon-phonon gas demonstrates how magnon-phonon scattering can significantly modify a formation of a Bose-Einstein condensate of an ensemble of magnons, providing a novel condensation phenomenon with a spontaneous accumulation of hybrid magnetoelastic bosonic quasiparticles~\cite{PhysRevLett.118.237201}.

Previous studies exemplifies how the interaction between magnons and vibrational lattice modes turned into a hot topic of research, considered as a powerful method either for spin control, as well as potential use as transduction from magnon signals to electrons~\cite{doi:10.1063/10.0000872}. Still on the promising character of the magnonics, we observe the recent advent of advanced materials exhibiting high-frequency magnons, which have been impelled the development of a new class of ultrafast spintronic devices~\cite{Jin2018,Qin2015,PhysRevLett.123.257202,PhysRevB.99.184439}. Such studies have reported terahertz magnons in the 2D Ising honeycomb ferromagnet CrI$_3$~\cite{Jin2018}, ultrathin film of iron-palladium alloys~\cite{Qin2015}, layered iron-cobalt magnonic crystals~\cite{PhysRevLett.123.257202} and noncollinear magnetic bilayers~\cite{PhysRevB.99.184439}. As we consider all the previously aspects, we are faced with the question: How does magnetic excitation behave under different time scales of the magnon and the lattice vibrations? In fact, the magnon excitation in a magnetoelastic lattice has an interdependent relaxation mechanism, and the formation of the magnon-polaron lacks a greater understanding. What would be the consequences of a spin transport as fast as the lattice dynamics? In order to answer these questions, we offer a systematic investigation of a quantum Heisenberg model, in which each spin of a one-dimensional lattice exhibits a vibrational degree of freedom around its equilibrium position. Such character is described by a standard Hamiltonian of coupled harmonic oscillators, whose magnetoelastic coupling is described by an exchange interaction which depends on the lattice deformations. By considering an intrinsic anisotropy mediated magnetoelastic coupling, we explore the Heisenberg spin-spin coupling within a XXZ framework, whose results demonstrate an original method for achieving a hybridized state referred as magnon-polaron. Such state is characterized by high cooperativity between the underlying excitations, in which a traveling or stationary formation depends on the magnetoelastic interaction. We reveal the critical amount of the magnon-lattice interaction ($\chi_c$) necessary to emergence of the static magnon-polaron quasi-particle. Bellow the critical magnetoelastic interaction, the magnon-polaron excitation develops two-fronts that propagates with constant velocity, whose spatial matching of their wave distributions exhibits a selection of particular modes. Their velocities continuously decreases with the power-law dependence as the magnon-lattice interaction grows. By exploring the ratio between the characteristic time scales of the magnon and the vibrational lattice, we unveil a limiting value of the critical magnetoelastic coupling, which is achieved as the magnon dynamics becomes much slower than the lattice dynamics.

\section{Model and formalism}
\label{sec:model}

\begin{figure}[!t]
    \centering
        \includegraphics[scale=0.75,clip]{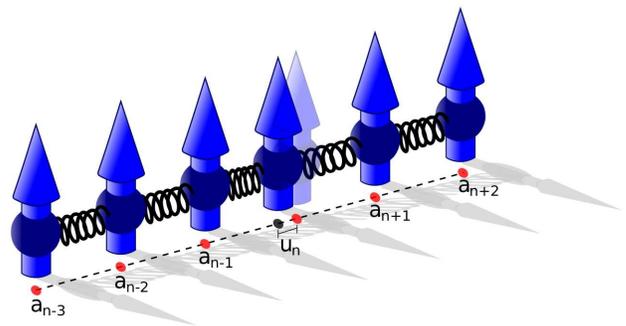}\\
    \caption{Pedagogical representation of the unidimensional spin lattice with effective harmonic springs coupling nearest-neighbor spins. The exchange interaction along the field direction ($J_{n,+1}^Z$)  depends linearly on the distance between  spins. Thus, the magnetic excitation contributes to the emergence of vibrational components, whose simultaneity and interaction mediated by the magnetoelastic coupling excites the magnon-polaron formation.}
   \label{fig1}
\end{figure}

The problem consist in analyzing  a one-dimensional magnetoelastic lattice, in which spins $1/2$ are located at lattice sites~(see fig~\ref{fig1}). We consider a crystal elastically isotropic, described by identical oscillators distributed along the lattice sites, which are ruled by a nearest-neighbor elastic coupling. Thus, the total Hamiltonian comprises magnetic and vibrational components
\begin{equation}
H = H_{mag} + H_{latt},
\end{equation}
with the vibrational contribution of system $\mathcal{H}_{latt} $ given by
\begin{eqnarray}
    \mathcal{H}_{latt} =
       \sum_{n} {{p_{n}^{2}}\over{2M}} 
          + {{\kappa}\over{2}} (u_{n+1}-u_{n})^{2}.
\end{eqnarray}
Here, $M$ represents the mass of the ions and $\kappa$ is the effective  spring's constant. Further, $p_{n}=M\dot{u}_{n}$ describes the conjugated momentum for the $n$-th spin. By considering the spin lattice along the $x$ axis, we parameterize the Hamiltonian in terms of the displacement $u_n=a'_n-a_n$, with $a'_n$ and $a_n$ denoting the respective position and equilibrium position of ion $n$. We consider $\omega \hbar << k_{B}T$, in which $\omega=\sqrt{\kappa/M}$, $k_{B}$ is the Boltzmann constant and $T$ the temperature. In this framework, the lattice dynamics can be treated within the classical mechanics formalism.

The magnetic component is governed by a quantum Heisenberg Hamiltonian, in which the spin-spin interactions are described by a nearest-neighbor exchange. We study the XXZ model
\begin{eqnarray}
    \begin{tabular}{rl}
         $\displaystyle \mathcal{H}_{mag} =$ & $\displaystyle E_0 + g\mu_{B}NH$ \\
          & $\displaystyle + \sum_{n} 2\hbar S(J_{n,n+1}^{~~Z}+J_{n,n-1}^{~~Z})c_{n}^{\dagger}c_{n}$ \\
         & $\displaystyle - 2\hbar SJ_{n,n+1}^{~XY}c_{n+1}^{\dagger}c_{n}
         - 2\hbar SJ_{n,n-1}^{~XY}c_{n-1}^{\dagger}c_{n}$,
   \end{tabular}
\end{eqnarray}
with $S = \hbar/2$, $J_{n,m}^{~Z} $ denoting the exchange anisotropic component between the $n$-th and $m$-th spins, and $J_{n, m}^{~ XY}$ describing the exchange coupling components in the XY plane.  The ground-state energy in the presence of a uniform external magnetic field is given by $E_0=-g\mu_{B} NHS-S^{2}\sum_{n}J_{n, n\pm 1}^{~~ Z} $. Here, we will focus on the propagation of magnetic excitation. Thus,  $c_{n}^{\dagger}$ and $c_{n}$ are respectively  the  creation  and annihilation operators at the $n$-th site.  Whenever the creation operator is applied to the ground-state, it leads to the excited state with the spin at site $n$ flipped. Furthermore, we consider an intrinsic anisotropy-mediated magnetoelastic coupling, such that the spin-spin coupling at the ``Z" direction (i.e the direction of external magnetic field) depends on the effective displacements between neighboring spins. We assume a regime of small amplitude oscillations described by
\begin{eqnarray}
        J_{n,n+1}^{~~Z}&=&J_{0}+\alpha(u_{n+1}-u_{n}) \nonumber \\
        J_{n,n+1}^{~XY}&=&J_{0},
\end{eqnarray}
where $\alpha$ denotes the effective spin-lattice coupling affecting the longitudinal spin-spin interactions. Such framework gives a ground-state energy ($ E_0 = -g \mu_{B}NHS-S^{2}NJ_{0}$) independent from the vibrational modes of the spin chain.

We analyze two key parameters: the effective coupling between the magnetic properties and the vibrational modes (i.e. the magnon-lattice interaction) $\chi = \hbar^2 \alpha ^2 /J_{0}\kappa $; and the ratio of the characteristic times scales of magnon ($ t_m = 1 / \hbar J_0$) and ionic chain ($ t_l = 1 / \omega $), written as $\tau= t_m / t_l$. By employing a normalized spin position $ x_ {n} = \sqrt{\kappa/\hbar^2 J_0}u_{n}$, the set of equations that describe the dynamics of the magnon and the lattice vibrations can be written respectively as  
\begin{eqnarray}
    i t_m \dot{\psi}_{n} = 2\psi_{n}-\psi_{n+1}-\psi_{n-1}-\sqrt{\chi}(x_{n+1}-x_{n-1})\psi_{n}~\nonumber
\end{eqnarray}    
and
\begin{eqnarray}
    {{t_m^2}\over{\tau^2}} \ddot{x}_{n} = x_{n+1}+x_{n-1}-2x_{n}-\sqrt{\chi}(|\psi_{n-1}|^2-|\psi_{n+1}|^2).\nonumber\\
    \label{mag}
\end{eqnarray}
The above set differential equations was solved by using a standard Runge-Kutta method with a time step small enough to keep the wave function norm conservation ($|1-\sum_n |\psi_n |^2| \leq 10^{-12}$ ) along the entire time interval considered. We concentrate our study by considering the initial state as single spin flip fully localized and centered at rest in static lattice center ($n = 0$ will be taken as the center of chain). Furthermore, we consider the characteristic time scale $t_{m} $ as the relevant time unit. We explore the regime with $\tau>1$ i.e   the lattice dynamics is faster than the magnon propagation. Although this regime is customary in ferromagnetic insulators, we consider the recent advent of advanced materials exhibiting high-frequency terahertz magnons~\cite{Jin2018,Qin2015,PhysRevLett.123.257202,PhysRevB.99.184439} for exploring regimes where the magnon dynamics is as fast as the lattice dynamics. Through the above-described procedure, we computed typical quantities able to bring information about the wave-packet time evolution, as will be detailed below.

\section{Results}

\begin{figure}[!t]
    \centering
        \includegraphics[trim=0mm 17mm 31mm 28mm, clip, height=3.9cm]{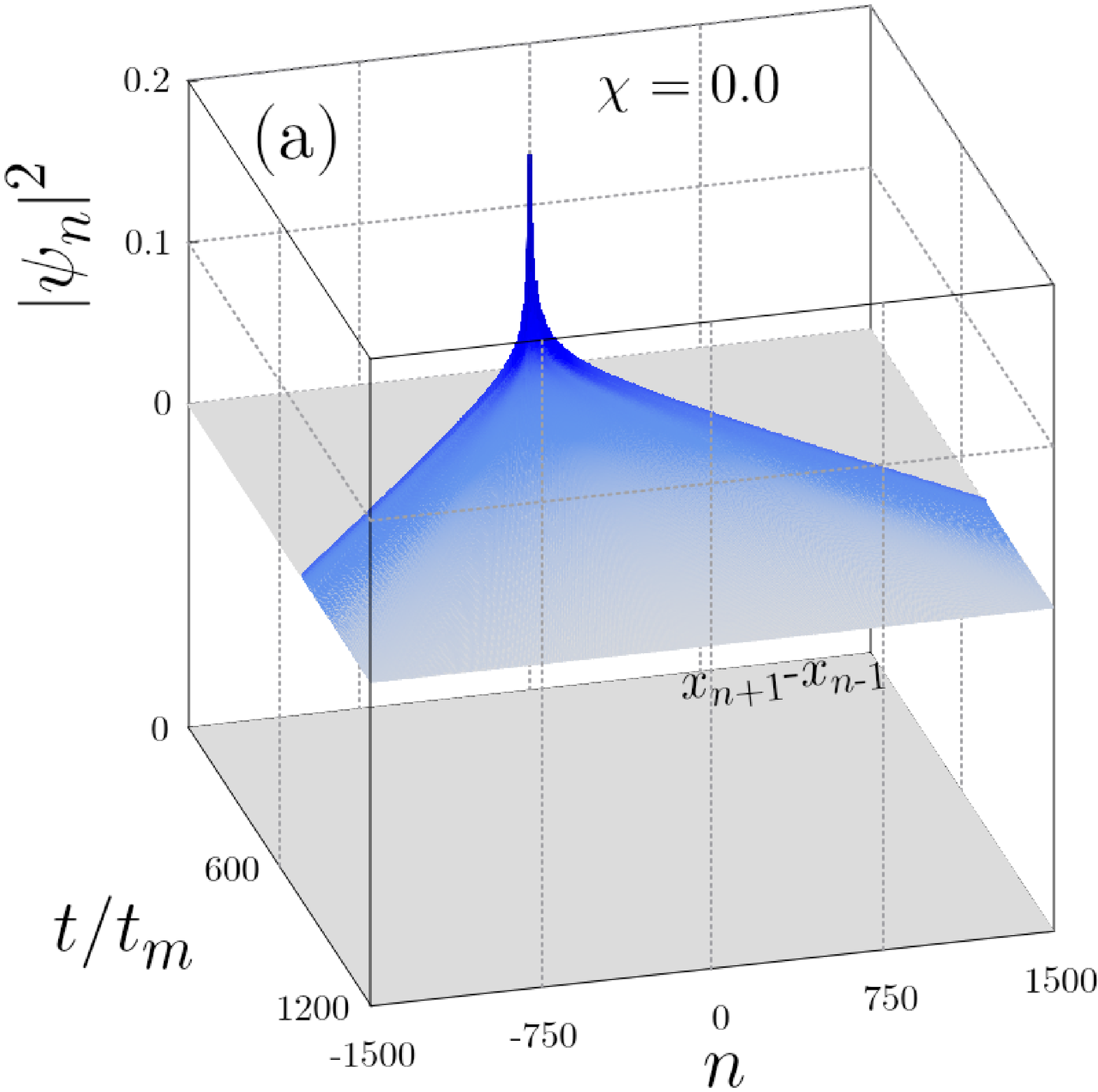}
    \includegraphics[trim=0mm 17mm 31mm 28mm, clip, height=3.9cm]{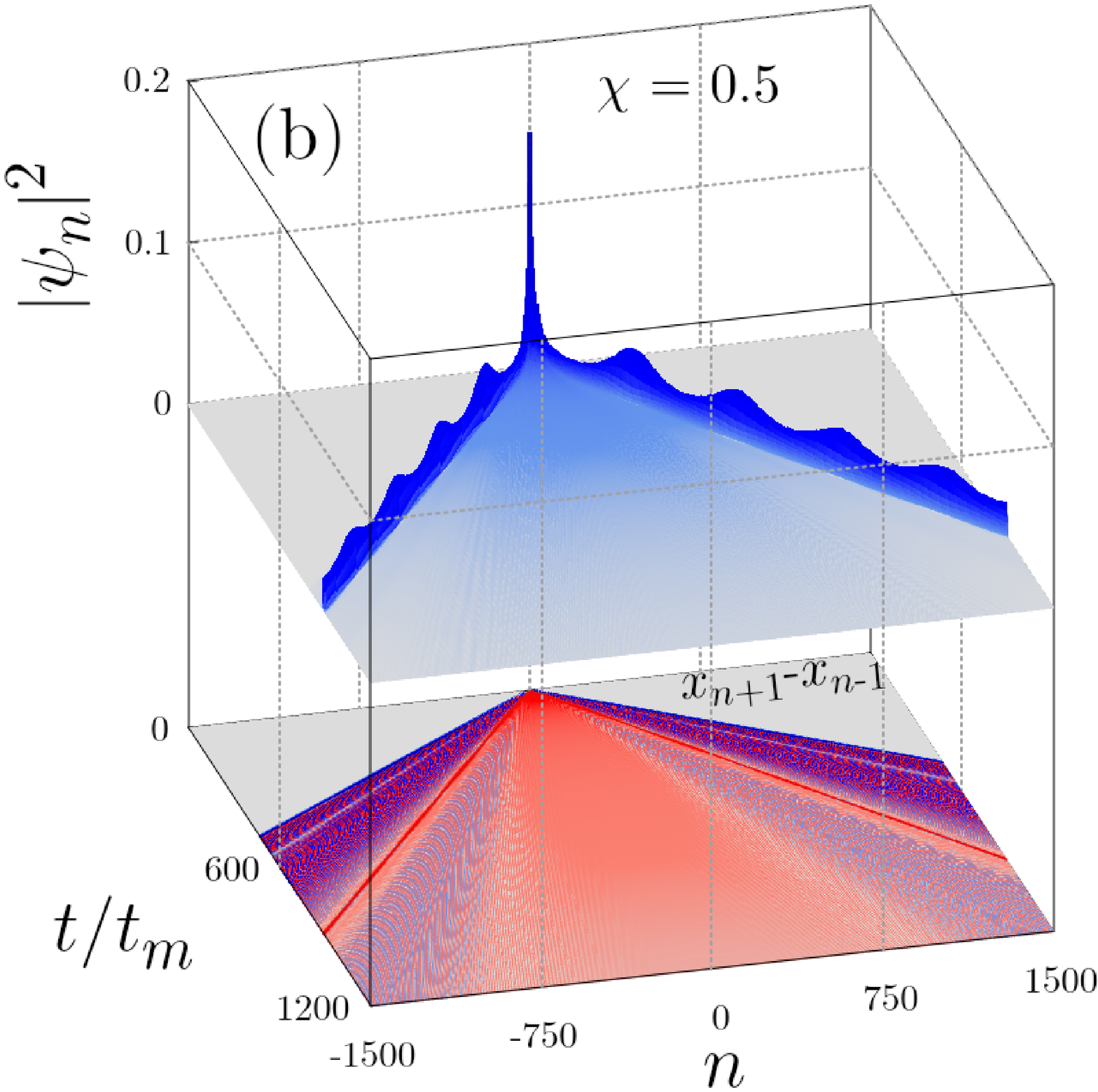}~~~~~~~\\
    \includegraphics[trim=0mm 17mm 31mm 28mm, clip, height=3.9cm]{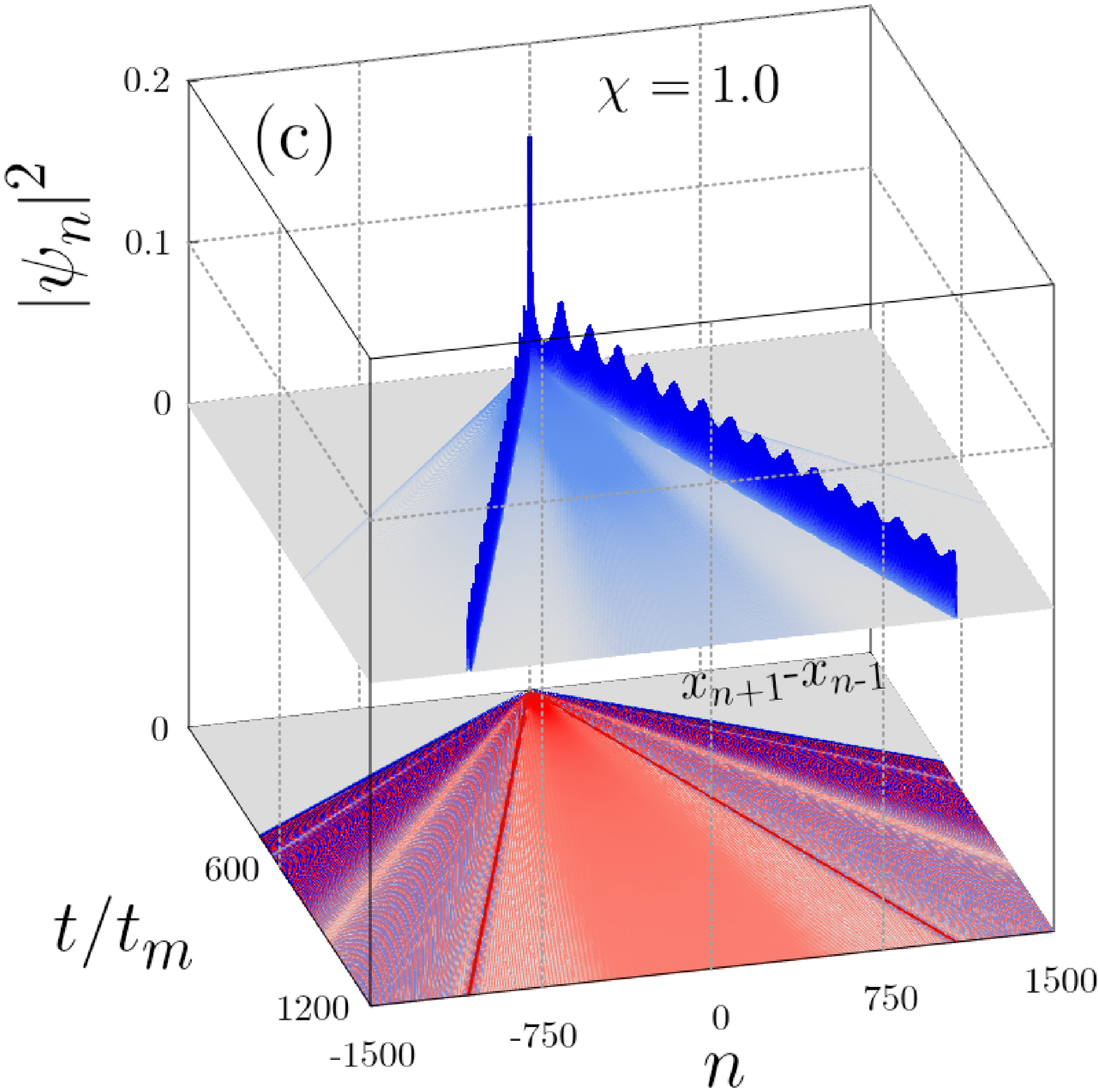}
    \includegraphics[trim=0mm 17mm 31mm 28mm, clip, height=3.9cm]{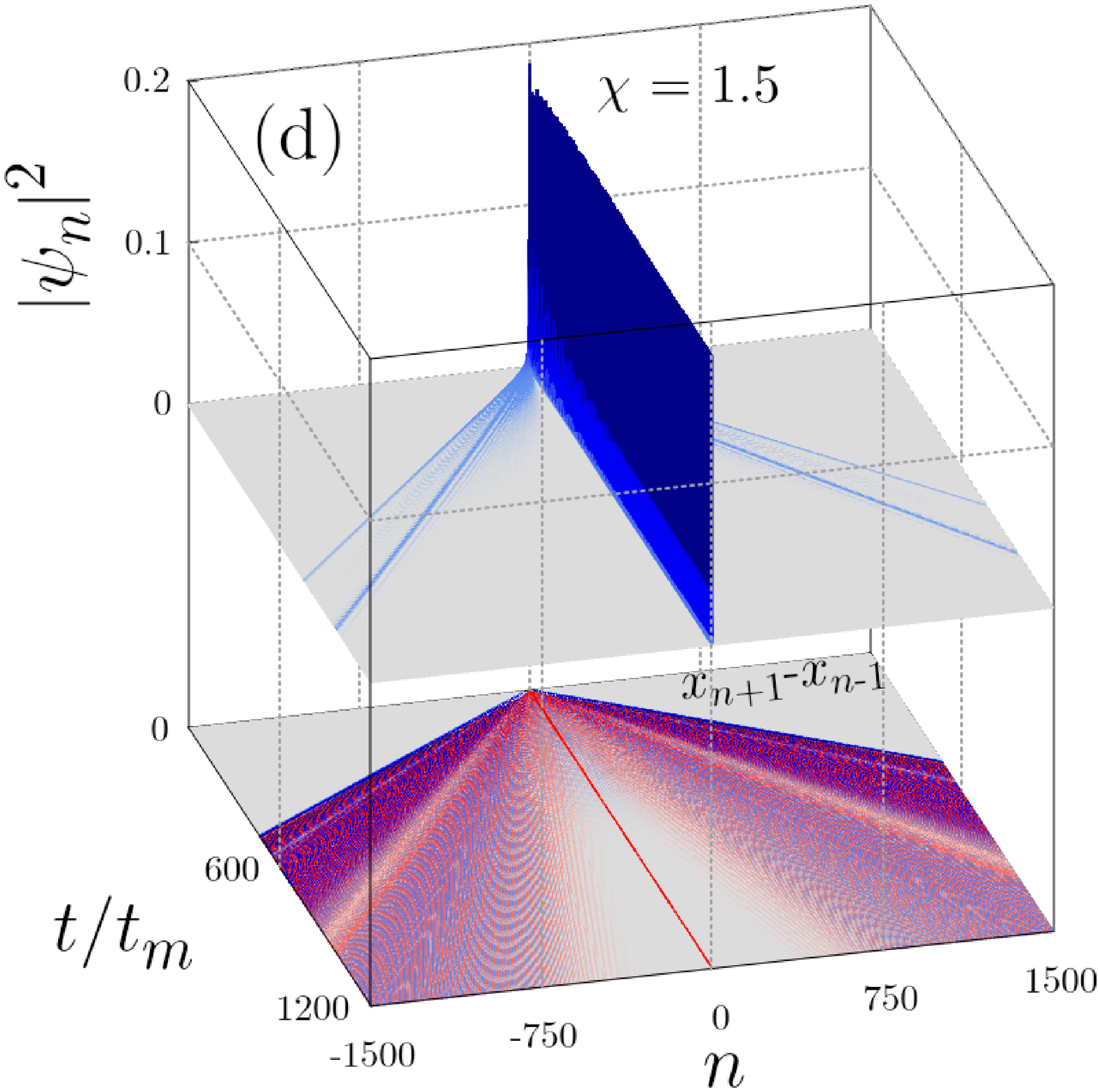}
    \caption{Time evolution of the magnon wave-function $ | \psi_ {n} | ^ 2 $ versus time and $n$ ($n=0$ represents the center of chain). The local  deformation   $x_{n + 1} -x_ {n-1}$ of chain is also investigated.  Calculations were done considering $ \tau = 10 ^ {0.5} $. (a) In the absence of magnon-lattice coupling ($ \chi = 0.0 $) the chain remains static and the magnon propagates ballisticaly along the chain  (b-c) considering $\chi=0.5$ and $1$ the coupling between lattice and spin-waves promotes the appearance of a solitonic like mode traveling along the lattice; (d) for $\chi=1.5$ the wave-packet remains trapped around the center of chain.  }
    \label{fig2}
\end{figure}

Using the numerical method described above, we start by examining the time evolution of initially fully localized magnon wave-packet at the center of a lattice initially 
in rest ($x_n=0$ and $\dot{x}_n=0$), with  $ \tau = 10^{0.5} $. In Figure \ref{fig2} we plot the time evolution of the wave-function profile $|\psi_n|^2$ and its respective lattice deformation $x_n-x_{n-1}$ for some representative values of magnetoelastic coupling $\chi$. In the absence of magnon-lattice interaction ($\chi = 0 $), we observe the magnon wave-function spreading ballistically over the entire lattice, that remains static. The scenario is significantly modified when we consider the interaction between magnon and lattice. For weak magnetoelastic coupling, breathing magnon modes emerge and propagate with constant velocity, while a fraction of the magnon wave radiates through the lattice (see Figs.~\ref{fig2}b-c). By following its respective lattice deformation, we observe signatures of the magnon-polaron formation, which explains the non-dispersive profile of the magnon wave-function. Magnon-polaron modes become slower as we increase the magnetoelastic coupling. A strong enough coupling induces a significant fraction of the magnetic excitation to remains trapped around its initial location (see Fig.~\ref{fig2}d). Such behavior is also characterized by a spatial matching between the spin-mode and lattice deformations distribution. Thus, we observe a high cooperation between the underlying excitations as a signature of the hybridized magnon-polaron state.

In order to better understand this rich set of dynamical profiles, we explore the participation function for the magnon
\begin{eqnarray}
 \xi (t) =  \sum_{n}1/|\psi_{n}(t)|^4,
\label{pp1}
\end{eqnarray}
and the lattice vibrations
\begin{eqnarray}
 \Xi = {{\left(\sum_{n}(x_{n+1}-x_{n})^2+{{\dot{x}_{n}^2}\over{\tau^2}}\right)^{2}}\over{\sum_{n}\left[{{(x_{n+1}-x_{n})^2}\over{2}}+{{(x_{n}-x_{n-1})^2}\over{2}}+{{\dot{x}_{n}^2}\over{\tau^2}}\right]^2}}.
\label{pp2}
\end{eqnarray}
\begin{figure}[t]
    \centering
    \includegraphics[clip,scale=0.45]{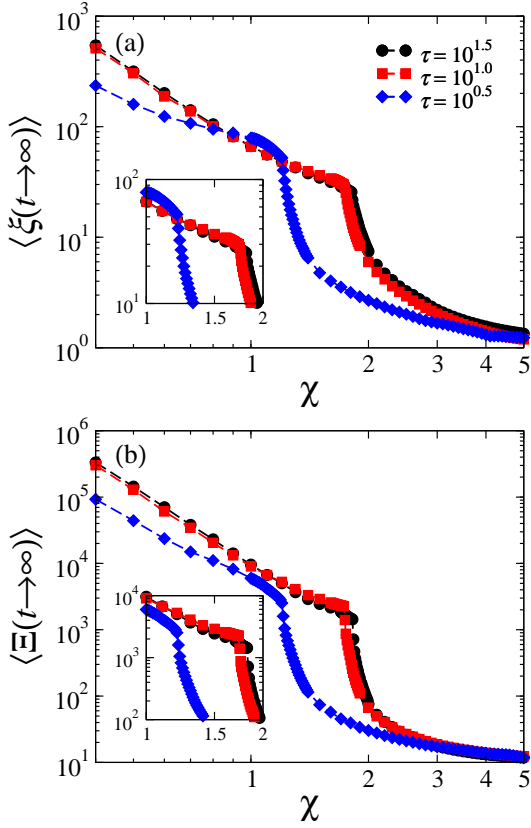}\\
    \caption{Average participation function for both magnon and lattice vibrations at the long time limit versus $\chi$, exploring $\tau=10^{0.5}$ up to $10^{1.5}$. Data suggest a kink singularity developed at $\chi_c$, above which the magnon-polaron formation stays stationary. Furthermore, such critical magnetoelastic coupling $\chi_c$ changes with different $\tau$.}
    \label{fig3}
\end{figure}
Such traditional quantities provide, respectively, an estimate of the number of lattice sites over which the magnon wave-function is spreading at time $t$, and the number of disturbed lattice sites at time $t$. Their scaling behavior can be used to distinguish the different dynamical regimes. The asymptotic participation function becomes size-independent for localized wave packets. On the other hand, $\langle \xi (t \to \infty) \rangle \propto N$ and $\langle \Xi (t \to \infty) \rangle \propto N$ corresponds to the regime where the magnon wave-packet and the lattice vibrations are uniformly distributed over the lattice. In fig. \ref{fig3} we compute the long-time behavior of the participation function for the magnon [$\langle \xi (t \to \infty) \rangle$] and the lattice dynamics [$\langle \Xi (t \to \infty) \rangle$] versus the effective magnetoelastic coupling $\chi$. Here, we explore the time scales of magnon and lattice vibrations. Calculations were done for $\tau=10^{0.5}$ up to $10^{1.5}$. Magnon and lattice vibrations decrease the propagation as the magnetoelastic coupling increases. This aspect corroborates the previous scenario of magnon-polaron formation, characterized by non-dispersive modes of spin and lattice vibrations exhibiting a constant velocity that decays as the magnetoelastic coupling increases. We further note an emergence of a kink singularity, which reveals an abrupt decreasing at participation functions as $\chi$ increases even more. Such behavior signals the critical point that establishes the beginning of the stationary regime, corroborated by full agreement exhibited between the asymptotic dynamics of the magnon and the lattice vibrations. Furthermore, the long-time participation function is vanishingly small as the magnetoelastic coupling increases even more, i.e. the degree of trapping is enhanced. The critical magnetoelastic coupling increases as the $\tau=t_ m/t_L$ grows, a consequence of the propagation of vibrational modes.


\begin{figure}[t]
    \centering
    \includegraphics[clip,scale=0.39]{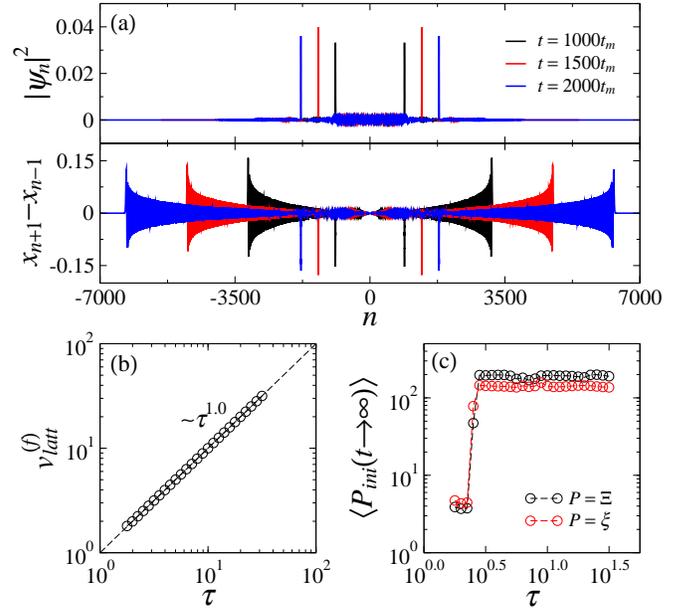}
    \caption{With $\chi=1.0$ and (a) $\tau=10^{0.5}$, the spatial profile of the magnon wave-function and lattice deformations describe the spatial matching and bound dynamics of the magnon-polaron formation, while (b) the wave-front velocity of the lattice vibrations versus $\tau$ shows a linear increasing with $\tau$. (c) The asymptotic participation functions around the initial magnetic excitation reveals a more pronounced spreading as $\tau$ increases. Thus, we observe as the lattice vibrations contribute to the magnon trapping and the consequent formation of stationary magnon-polaron.}
    \label{fig4}
\end{figure}

In Fig.~\ref{fig4}a we display snapshots of the magnon wave-packet and the corresponding lattice deformations for $\chi=1.0$ and $\tau=10^{0.5}$. Besides magnon and lattice deformations exhibiting non-dispersive modes with perfect spatial match, the lattice deformations are spreading over the lattice by developing wave-fronts. With the magnon approximately as fast as the lattice dynamics, disturbances originating from the lattice wave-fronts that extend along the tails inhibit the propagation of the magnon wave-packet. Thus, a smaller coupling between magnetic and mechanical components is required to the stationary formation of the magnon-polaron. This character is better understood when looking at Fig.~\ref{fig4}b-c, where we remain with $\chi=1.0$, but we explore a range of $\tau$. Fig.~\ref{fig4}b shows the wave-front velocity of the lattice deformations exhibiting a linear growth with $\tau$.  Fig.~\ref{fig4}c displays the asymptotic participation functions around the initial site of magnon excitation [$n_0-100 \leq n \leq n_0 + 100$]. A small ratio between characteristic time scales of magnon and lattice vibration favors the stationary magnon-polaron formation, described by  $\langle P_{ini}(t \to \infty) \rangle \approx 1$. With the magnon spreading slowly enough, the magnetoelastic coupling is unable to establish stationary formation.

\begin{figure}[t]
    \centering
     \includegraphics[clip,scale=0.44]{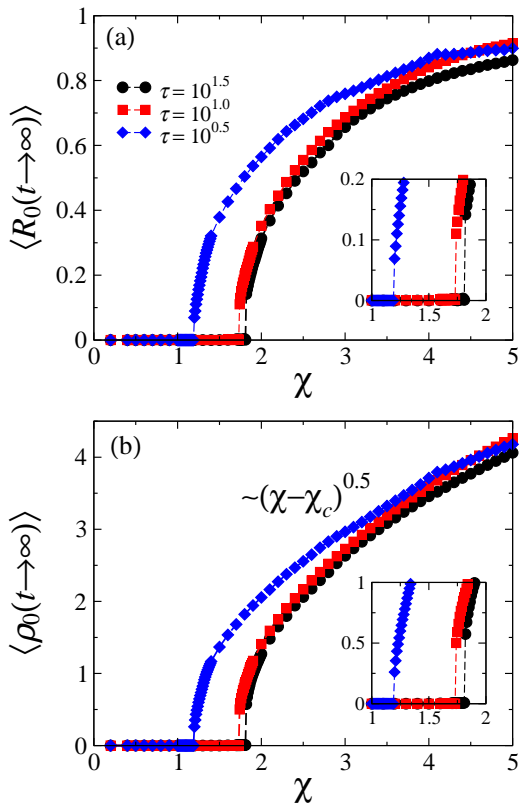}
    \caption{The asymptotic return probabilities of the (a) magnon and (b) the lattice vibrations exhibit clear signatures of a phase transition between the traveling  and stationary regime of magnon-polaron formation, which corroborates the results at the Fig.~\ref{fig3}. Above the critical magnetoelastic coupling $\chi_c$, magnon and lattice vibrations become significantly clustered around the site of the initial magnetic excitation.}
    \label{fig5}
\end{figure}

\begin{figure}[t]
    \centering
       \includegraphics[clip,scale=0.44]{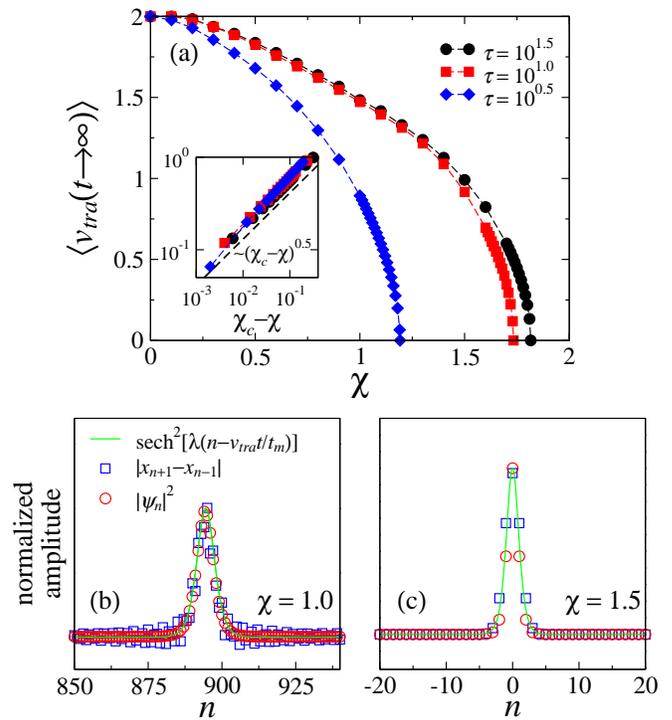}
    \caption{Dynamics of the traveling formation of magnon-polaron, with (a) the velocity versus $\chi$ and its respective scaling analysis. The best power-law fit provides $v \propto ( \chi_c-\chi)^{1/2}$. The spatial profile of the magnon wave-function and the matching lattice deformation for the magnon-polaron state, whether (b) traveling or (c) stationary, corroborates the characteristic spatial profile from the breathing bright solitons (fitting curves in solid lines).}
    \label{fig6}
\end{figure}

We also explore the asymptotic regime of the return probabilities for the magnon and the lattice deformation:
\begin{equation}
R_{0}(t) =|\psi_{n=0}(t)|^{2} \hspace{.3cm} \mbox{and} \hspace{.3cm} \rho_{0}(t) = |x_{1}(t)-x_{-1}(t)|.
\end{equation}
Such measures offer the probability of finding the magnon wave-packet or the lattice deformations at the position corresponding to the initial magnetic excitation. Thus, their scaling behaviors can also be used to distinguish between localized and delocalized wave packets in the long-time regime, with $R_0(t\to\infty)\to 0$ and $\rho_0(t\to\infty)\to 0$ connoting the magnon wave-function and the lattice deformations escaping from its initial position, respectively. On the other hand, the return probability saturates at a finite value for a stationary regime of the magnon-polaron. We observe in Fig.~\ref{fig5} the asymptotic behavior of both return probabilities corroborating the previous results. For $\chi<\chi_c$, the magnon and the lattice deformations exhibit a vanishingly small return probability, which confirms a predominant spreading through the lattice. Above a critical magnetoelastic coupling, the asymptotic return probabilities become significantly larger than $1/N$. Allied to the monotonic growth of both return probabilities, such behavior reinforces the emerging self-trapping regime described earlier. Both quantities exhibit a trend $(\chi- \chi_{c})^{0.5}$, as well as confirm a dependence on the parameter $\tau$ (see inset).

\begin{figure}[t]
    \centering
        \includegraphics[clip,scale=0.47]{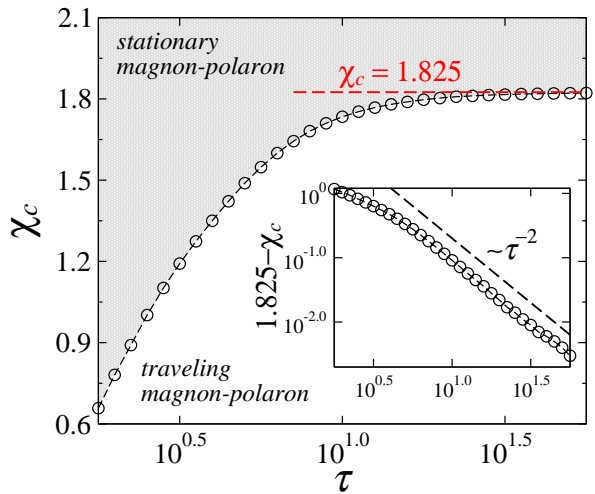}
    \caption{Plot of $\chi_c$ versus $\tau$ phase diagram. Corroborating previous results, the magnetoelastic coupling necessary to the stationary formation of magnon-polaron increases as the $\tau$ growth. However, such this behavior is restricted to the small enough $ \tau $ regime. The further increasing in the ratio of characteristic time-scales unveils a limit value $\chi_c^{max}$, above which the critical coupling becomes indifferent to the $\tau$.  The best fitting is achieved for $\chi_c^{max} = 1.825$, as well as $(\chi_c^{max}-\chi_c) \propto \tau^{-2}$ (see inset).}
    \label{fig7}
\end{figure}

The magnon-polaron formation and its threshold between traveling and stationary regimes have also been identified by exploring the magnon-polaron velocity. For the Fig.~\ref{fig6}a, we measure the mean velocity of the magnon-polaron modes at the long-time regime [$\langle v_{tra}(t \to \infty) \rangle$] and investigate its relationship with $\chi$. Data show the velocity decreasing as the effective magnetoelastic coupling increases, up to the threshold at which vanishes ($\chi\geq\chi_c$).  Besides the critical magnetoelastic coupling showing full agreement with the measures of participation function (see Fig.~\ref{fig3}) and return probability (see Fig.~\ref{fig5}), 
the traveling velocity of the magnon-polaron decreases with $\langle v_{tra} (t \to \infty) \rangle \propto (\chi_{c} - \chi)^{0.5}$, the same exponent exhibited by the asymptotic return probability after the critical point. 

In order to better understand such hybridized excitation of magnon-polaron, in Fig.~\ref{fig6}b-c we show profiles of the magnon wave-function and the matching lattice deformation in a snapshot achieved during the time-evolution for a system ruled by $\tau=10^{0.5}$ and $\chi=1.0, 1.5$. The magnon wave-packet follows the standard solitonic profile $\mathrm{sech}^2[\lambda(n + vt / t_ {m})]$)~\cite{holstein1959a,holstein1959b}, either for the traveling (see Fig.~\ref{fig6}b) and the stationary excitations (see Fig.~\ref{fig6}c). Such magnetic components are bound to a lattice structural kink that also exhibits the well-known breathing bright soliton-like spatial profile.

The consequences of different time scales of the magnetic and vibrational components are exhibited in Fig.~\ref{fig7}, in which we extend our numerical experiments in order to offer $\chi_c$ versus $\tau$ diagram. For greater accuracy, data have been computed by analyzing the participation function, the return probability, as well as the magnon-polaron velocity. Systems in which the dynamics of magnons is comparable to the lattice dynamics show an increase in the critical magnetoelastic coupling $ \chi_c $ as $\tau $ increases. However, when considering systems with an ever slower magnon dynamics, this behavior leads monotonically $\chi_c $ to a limit value [$ \chi_c \approx 1.82(1)$], above which the increase in the $ t_m / t_L $ ratio becomes indifferent.  By analyzing the critical magnetoelastic coupling versus $\tau$, the best fitting provides $\chi_c^{max}\approx 1.825(5)$ and $(\chi_c^{max}-\chi_c) \propto \tau^{-2}$ (see inset).

\section{Summary}
\label{sec:sum}

In this work we study how the lattice dynamics influence the dynamics of initially localized one-magnon excitations. We consider a quantum anisotropic Heisenberg ferromagnetic chain, in which the spins $1/2$ belongs to a chain of coupled Harmonic oscillators. The magnon-elastic coupling was introduced by considering the longitudinal spin-spin exchange coupling as a linear function of the effective displacement between nearest-neighbor spins. By exploring the numerical solutions of the set of equations that govern the system, our results exhibit a framework for obtaining the magnon-polaron formation, whose features are ruled by the effective magnetoelastic coupling, as well as lattice and magnon characteristic time scales. The features of this hybridized state, with high-cooperation between magnetic and mechanical components, are closely related to magnetoelastic coupling. Weak enough couplings promote the formation of traveling magnon-polarons, whose velocity depends on the strength of the magnetoelastic coupling. When analyzing stronger couplings, we are faced with a localization scenario, in which the magnon-polaron formation remains trapped around the position of the initial magnetic excitation. This scenario was described by different physical quantities, such as participation function, probability of return, as well as the velocity of traveling formation of magnon-polaron. The current numerical results provide accurate estimates of the critical point and respective singularities of the relevant quantities associated with the transition to the stationary regime. In addition, we reveal the magnon-polaron formation exhibiting a soliton-like profile. The critical magnetoelastic coupling that separates the stationary and traveling regimes is related to how much the magnon is slower than the lattice vibrations. It increases proportionally with the time scale of the magnon until a limit value is reached, above which the magnon dynamics no longer interferes at the critical magnetoelastic coupling capable of inducing the stationary magnon-polaron. Our study aims to contribute with the emergent development of a new class of ultra-fast spintronic devices, and the consequent applications of magnon-polaron. Considering that the present status of experiments with cold atoms trapped in optical lattices allows the study of Heisenberg XXZ and a fine control of a wide collection of anisotropies including the XX and XXX limits~\cite{naturedezem2020}, we believe in the feasibility of the scheme proposed here. To conclude, it would be interesting to have these results derived from an analytical framework, which would bring valuable new insights into the general dynamics involved in magnon-polaron formation. Future works that explore the soliton-like profile of the magnon-polaron formation, such as coherence properties and binary collisions, may contribute to a better understanding and applicability.

\section{Acknowledgments}
\label{sec:acknowl}

This work was partially supported by CNPq (Brazilian National Council for Scientific and Technological Development),
CAPES and FINEP (Federal Brazilian Agencies), as well as FAPEAL (Alagoas  State Agency). 

\bibliography{referencias} 

\end{document}